\documentclass[twocolumn,prl,showpacs,amsmath,amssymb]{revtex4}

\usepackage{graphicx}
\usepackage{dcolumn}
\usepackage{bm}
\usepackage{amssymb}

\begin{document}

\title{Crossed Andreev reflection-induced magnetoresistance}

\author{Francesco Giazotto}
\email{giazotto@sns.it}
\affiliation{NEST CNR-INFM and Scuola Normale Superiore, I-56126 Pisa, Italy}
\author{Fabio Taddei}
\email{taddei@sns.it}
\affiliation{NEST CNR-INFM and Scuola Normale Superiore, I-56126 Pisa, Italy}
\author{Rosario Fazio}
\affiliation{NEST CNR-INFM and Scuola Normale Superiore, I-56126 Pisa, Italy\\
and International School for Advanced Studies (SISSA), I-34014 Trieste, Italy}
\author{Fabio Beltram}
\affiliation{NEST CNR-INFM and Scuola Normale Superiore, I-56126 Pisa, Italy}


\begin{abstract}
We show that very large negative magnetoresistance can be obtained in magnetic trilayers  in a  current-in-plane geometry owing to the existence of \emph{crossed} Andreev reflection.
This spin-valve consists of a thin superconducting film sandwiched between two 
ferromagnetic layers whose magnetization is allowed to be either parallelly or 
antiparallelly aligned. For a suitable choice of structure parameters and 
nearly fully spin-polarized ferromagnets the magnetoresistance can exceed $-80\%$.
Our results are relevant for the design and implementation of spintronic devices exploiting ferromagnet-superconductor structures.
\end{abstract}


\maketitle

Giant Magneto Resistance (GMR) is the pronounced response in
the resistance of magnetic multilayers to an applied magnetic 
field~\cite{prinz98,ansermet98,wolf01,zutic04,baibich88}.
This phenomenon has prompted a very large interest
owing to its broad range of applications, spanning from magnetic recording 
to position sensor technology, and to the fundamental interest in spin-dependent 
effects~\cite{zutic04}.
A magnetic multilayer consists of an alternating sequence of ferromagnetic (F) and non-magnetic layers (N). 
The relative orientation of magnetic moments in the F 
layers can be driven from antiparallel (AP), in the absence of external field, 
to parallel (P), with a small (up to some hundreds of Oe)  magnetic field. GMR was 
originally demonstrated~\cite{baibich88} in Fe/Cr multilayers with current flowing 
parallel to the planes, the so-called current-in-plane (CIP) configuration. 
In the CIP measurement the magneto-resistance (MR) ratio, defined as the maximum relative change in resistance 
resulting from applying the external field, is typically around $10\%$ for a number
of layers of the order of $50-100$~\cite{baibich88}. These values
 can be increased up to $\sim 100\%$ in the case of current flow perpendicular to the multilayer plane (CPP configuration)~\cite{pratt91}. 

In this Letter we show that the limitations of the CIP configuration can be overcome by employing a \emph{superconductor} (S) in the non-magnetic portion of the multilayer. 
The use of superconductors in spintronics is not new. As a matter of fact,
superconductors were used already in the very first CPP experiment~\cite{pratt91} in order to minimize the extra 
resistance introduced in contacting the multilayered structure to the measuring apparatus.
The peculiar properties of FS structures have been studied for several years and this field has been recently reviewed in Ref.~\cite{fsreviews}.
\begin{figure}[t!]
\includegraphics[width=8cm,clip]{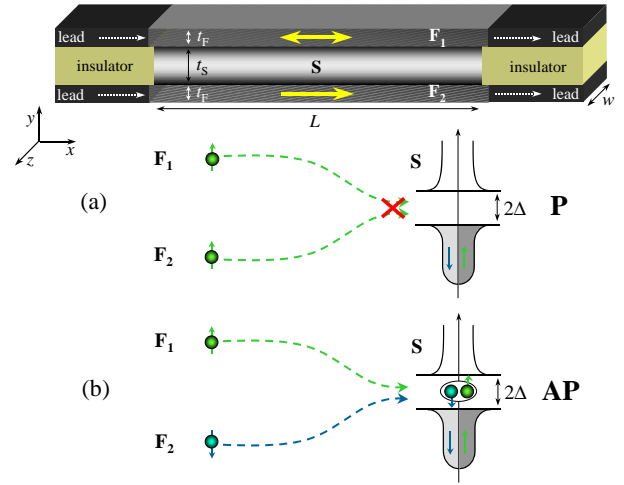}
\caption{(color online) Sketch of the FSF  spin-valve. A thin superconducting 
	film is sandwiched between two identical ferromagnetic layers whose 
	magnetizations (yellow arrows) can be aligned both in the parallel (P) 
	and antiparallel (AP) configuration.  An electric current (white dashed arrows) 
	is allowed to flow through the system parallel to the layers. 
	The schematic representation of the spin-valve effect for \emph{half-metallic} ferromagnets, showing the 
	diagrams of the superconducting density of states, is displayed in (a) and (b). 
	(a) In the P alignment, the lack of quasiparticles with  opposite spin hinders the
	condensation of two electrons injected from the ferromagnets in a Cooper pair 
	in S. As a consequence, the electric transport is confined within the F layers. 
	(b) In the AP configuration, two electrons with opposite spin injected from the F 
	layers  can form a Cooper pair within the superconductor thanks to crossed Andreev reflection, thus "shunting" the 
	current through the whole structure (see text).}
\label{fig1}
\end{figure}

The structure we envision (see Fig. 1) consists of two identical diffusive ferromagnetic layers (F$_1$ and 
F$_2$), of thickness $t_{\text{F}}$, separated by a (s-wave) superconducting layer of thickness 
$t_{\text{S}}$. The layers are assumed to be in good metallic contact and have length $L$ and width $w$. The magnetization of the two ferromagnets is allowed to be aligned either in  a parallel or an antiparallel configuration \cite{AFM}. The trilayer is connected to ferromagnetic leads separated by an 
insulating layer (light-yellow regions in Fig.~1) of the same thickness as the S layer. 
The magnetization of the upper F leads is equal to the one relative to layer F$_1$, 
and analogously for the lower F leads.
In the CIP configuration, charge transport in the system is dominated by \emph{crossed} Andreev reflection (CAR) leading to a dramatic enhancement of the 
magnetoresistance. CAR was analyzed in several papers~\cite{carth}, notably in 
relation to quantum information processing~\cite{entanglement}, and very recently it was 
observed experimentally in FS~\cite{carexp} and in NS~\cite{russo} structures. 
Here we emphasize its potential for spintronics.

Let us first describe qualitatively the principle of operation of the present spin-valve. For the sake of clarity, let us first consider a \emph{half-metallic} (i.e., with only one spin species) ferromagnet~\cite{coey} in good metallic contact with a S layer. Quasiparticles with energy below the superconductor gap can be transferred into the superconductor as Cooper pairs only through an Andreev reflection (AR) process \cite{andreev64}. 
The latter consists of a coherent scattering event in which a spin-up(down) 
electron-like quasiparticle, originating from the F layer, is retroreflected at 
the interface with the superconductor as a spin-down(up) hole-like quasiparticle 
into the ferromagnet. Since only quasiparticles (electron- and hole-like) of 
one spin type exist in the ferromagnet, no current can flow between 
the F and S layers \cite{beenakkerFS}. Similarly, in the case of the FSF 
trilayer in the P configuration (see Fig. 1(a)), the two F layers cannot transfer charge into the superconductor. Current is confined to the F layers and it consists of fully-polarized quasiparticles. If the S layer is thin enough quasiparticles can also tunnel through it (this will occur for $t_{\text{S}}$ values up to some superconductor coherence lengths ($\xi_0$)).
In the AP configuration (see Fig. 1(b)), each of the two F layers can contribute separately to the quasiparticle current through the structure just like in the P configuration. More importantly CAR does take place. In this case a Cooper pair is formed in the superconductor by a spin-up electron originating from the F$_1$ layer and a spin-down electron from the F$_2$ layer. In the AR language, this can be described as the transmission of a spin-up electron-like 
quasiparticle from one of the F layers to a spin-down hole-like quasiparticle in 
the other F layer. This is now possible since the quasiparticles involved belong to the majority spin species in each of the two layers. A charge current can therefore flow through the S layer as a \emph{supercurrent}, thereby shunting the conduction channels in the ferromagnets \cite{JK}. This contribution to the current will dominate at least when the structure is long enough and the quasiparticle contribution in the F layers becomes negligible (note that the conductance of each F layer in the diffusive regime is proportional to $\ell/L$, where $\ell\ll L$ is the mean free path).
As a result, one can expect the conductance $G_{\text{AP}}$ of the AP configuration to be much larger than the conductance $G_{\text{P}}$ of the P configuration. This can give rise to a large, \emph{negative} value of the MR ratio, defined as:
\begin{equation}
\text{MR}=\frac{G_{\text{P}}-G_{\text{AP}}}{G_{\text{P}}}.
\label{MR}
\end{equation}

A simple expression for the MR ratio for half-metallic ferromagnets 
in the diffusive regime can be derived as follows. In the P configuration, 
the conductance is approximately given by \cite{beenakkerFS}
\begin{equation}
G_\text{P}\simeq 2\frac{e^2}{h}\frac{\ell}{L}N_\uparrow, 
\label{gp}
\end{equation}
i.e., it is proportional to the number $N_\uparrow$ of open channels for spin-up 
electrons of each F layer, and inversely proportional to $L$.
In the AP configuration, the conductance can be roughly separated in two contributions. One ($G^{\ast}$), due to CAR, is virtually \emph{independent} of $L$. The other comes from the direct transmission of quasiparticles (proportional to $(2e^2/h)(\ell/L)N_\uparrow$):
\begin{equation}
G_\text{AP}\simeq G^{\ast}+\alpha \,2\frac{e^2}{h}\frac{\ell}{L}N_\uparrow,
\label{gap}
\end{equation}
with $\alpha$ being a numerical factor $\sim 1$.
As a result,
\begin{equation} 
\text{MR}\simeq 1-\alpha-G^{\ast}\frac{h}{2e^2} \frac{L}{\ell}\frac{1}{N_\uparrow}, 
\end{equation}
 negative and large for $L\gg\ell$. This is in contrast to what expected in a FNF trilayer, 
where the MR value is \emph{positive} \cite{baibich88} since the AP configuration yields a reduction of the structure conductance.  For non half-metallic ferromagnets, 
the charge current will still be dominated by CAR, but the effect will be reduced.

This qualitative understanding of the effect can be validated by a numerical calculation of the conductance, which was performed within the Landauer-B\"uttiker scattering approach.
In the presence of superconductivity, the zero-temperature and zero-bias conductance can be 
written as $G= G_{\uparrow}+G_{\downarrow}$~\cite{lambert93},
where
\begin{equation}
	G_{\sigma}=\frac{e^2}{h}\left[\mathcal{T}^{\sigma}+\mathcal{T}_{\text{a}}^{\sigma}+ 
	2\frac{\mathcal{R}_{\text{a}}^{\sigma} {\mathcal{R}^{\sigma}_{\text{a}}}'-\mathcal{T}^{\sigma}_{\text{a}} 
	{\mathcal{T}^{\sigma}_{\text{a}}}'}{\mathcal{R}^{\sigma}_{\text{a}}+{\mathcal{R}^{\sigma}_{\text{a}}}'+
	\mathcal{T}^{\sigma}_{\text{a}}+{\mathcal{T}^{\sigma}_{\text{a}}}'}\right]
\label{conductance2}
\end{equation}
is the spin-dependent conductance \cite{GTV}.
In Eq.~\ref{conductance2}, $\mathcal{T}^{\sigma}$ ($\mathcal{T}_{\text{a}}^{\sigma}$) is the spin-dependent normal (Andreev) transmission probability for quasi-particles injected from the left lead and arriving on the right lead, while $\mathcal{R}_{\text{a}}^{\sigma}$ is the Andreev 
reflection probability for quasi-particles injected from the left lead \cite{ARCAR}.
Similarly, ${\mathcal{T}_{\text{a}}^{\sigma}}'$ and ${\mathcal{R}_\text{a}^{\sigma}}'$ are the Andreev 
scattering probabilities for quasiparticles injected from the right lead. 
$e$ is the electron charge and $h$ is the Planck constant.
\begin{figure}[t!]
\includegraphics[width=\columnwidth,clip]{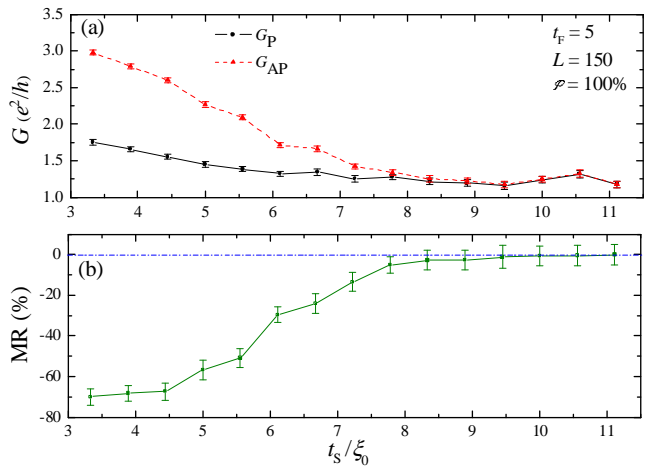}
\caption{(color online) (a) Conductance in the P (black circles) and AP (red triangles) 
	configurations versus $t_{\text{S}}$ with $t_{\text{F}}=5$. 
	(b) Resulting MR ratio. Data were obtained assuming $L=150$, $\mathcal{P}=100\%$, and
	 $U=8$ (see text). In (a) the error bars correspond to 
	the standard error over all disorder configurations. Lines are guides to the eye.}
\label{tS}
\end{figure}
The scattering amplitudes were evaluated numerically by making use of 
a recursive Green's function technique based on a tight-binding 
version~\cite{sanvito99} of the Bogoliubov-de Gennes equations
\begin{equation}
	\left( \begin{array}{cc} \mathcal{H} & \Delta \\ \Delta^* & -\mathcal{H}^*
	\end{array}\right)
	\left( \begin{array}{c} u\\v\end{array}\right)=E
	\left( \begin{array}{c} u\\v\end{array}\right) ,
\label{BdG}
\end{equation}
where $\mathcal{H}$ is the single-particle Hamiltonian, and $u$ ($v$) is the 
coherence factor for electron- (hole)-like excitations of energy $E$, 
measured from the condensate chemical potential $\mu$. Within the tight-binding 
description, $\mathcal{H}$ and $\Delta$ are matrices with elements 
$(\mathcal{H})_{ij}=\epsilon_i \delta_{ij}-\gamma\delta_{\{i,j\}}$ and 
$(\Delta)_{ij}=\Delta_i\delta_{ij}$, where $\epsilon_i$ is the on-site 
energy at site $i$, $\gamma$ is the hopping potential and $\Delta_i$ is the 
superconducting gap ( $\{...\}$ stand for first-nearest-neighbor sites).
In particular, $\epsilon_i=\epsilon_{\text{S}}$ in the S region, 
$\epsilon_i=\epsilon_{\text{I}}$ in the insulating barrier, and 
$\epsilon_i=\epsilon_{\text{F}}=\epsilon_{\text{S}}\mp h_{\text{exc}}$ 
in the F layers, $h_{\text{exc}}$ denoting the ferromagnetic exchange energy, 
with upper (lower) sign referring to majority (minority) spin species.
$\Delta_i$ is assumed to be constant and equal to zero-temperature gap ($\Delta_0$) 
in the S region, and zero everywhere else. Note that this is realistic when the S 
layer thickness is larger than $\xi_0$ \cite{you04}.
Furthermore, disorder due both to impurities and lattice imperfections is 
introduced by the Anderson model, i.e., by adding to each on-site energy a 
random number chosen in the range $[-U/2,U/2]$, being $U$ a fraction of the 
Fermi energy. In what follows we shall indicate energies in units of $\Delta_{0}$, 
and lengths in units of the lattice constant $a$ (of the order of the Fermi wavelength).

In order to analyze the behavior of conductances and MR as a function of the 
various parameters we used a two-dimensional (2D) model of the structure, i.e., 
we assumed a single lattice site in the $z$-direction (see Fig. 1).
In our calculations the tight-binding parameters were chosen to describe 
metallic materials: $\epsilon_{\text{S}}=20$, $\epsilon_{\text{I}}=10^3$, 
$\gamma=10$, so that $\xi_0 =(2a/\pi)\sqrt{4(\gamma /\Delta_{0})^2-\epsilon_{0}\gamma/\Delta_{0}^2}=9.0$.
We set $U=8$ and $L=150$, so that the F layers are in the diffusive regime. 
To avoid a self-consistent calculation of the superconducting gap, 
we limited our analysis to values of $t_{\text{S}}\geq 30$ 
(corresponding to $\simeq 3.3\xi_0$) \cite{you04,self}.
In addition, the conductance was calculated performing an ensemble average 
over $100$ realizations of disorder. 

\begin{figure}[t!]
\includegraphics[width=\columnwidth,clip]{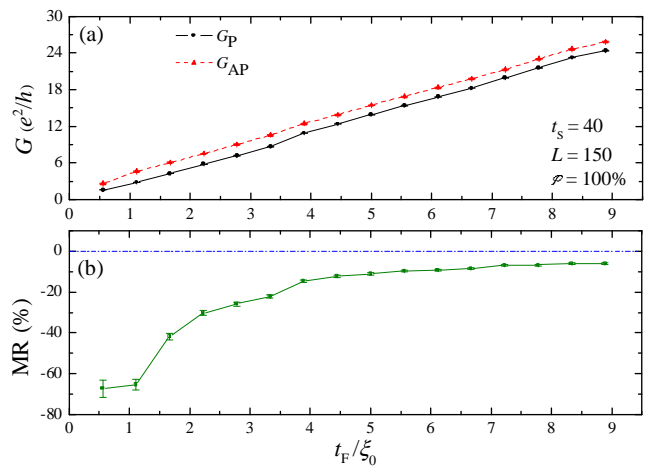}
\caption{(color online) (a) Conductance in the P (black circles) and 
	AP (red triangles) configurations versus $t_{\text{F}}$ with 
	$t_{\text{S}}=40$. (b) Resulting MR ratio. The same parameters 
	as for Fig.~\ref{tS} were used.}
\label{tF}
\end{figure}
The conductance and  MR dependence on S layer thickness is shown in 
Fig.~\ref{tS}. Here we chose the ferromagnetic thickness 
 $t_{\text{F}}=5$ and $h_{\text{exc}}=20$ (the mean free path turns out to be $\ell \simeq 21$). For this latter value the 
 ferromagnet polarization ($\mathcal {P}$)~\cite{polarizzazione} 
is equal to $100\%$.
 Figure~\ref{tS}(a) shows that in the P configuration 
the conductance $G_{\text{P}}$ is initially slightly decreasing and roughly constant for $t_\text{S}\ge 5.5\xi_{0}$.
This is due to the fact that quasiparticles 
in the two F layers (for large enough $t_{\text{S}}$ values $\ge 5.5\xi_{0}$) are decoupled, but some direct tunneling can occur through thinner S layers.
In the AP configuration, the conductance $G_{\text{AP}}$ decreases 
until the value $t_{\text{S}}\simeq 8.5\xi_{0}$ is reached, and thereafter remains almost constant.
Such a behavior is expected since, on the one hand, for $t_{\text{S}}$ of 
the order of some $\xi_0$ the conductance is dominated by the supercurrent 
(mediated by CAR between the F$_1$ and F$_2$ layers). 
On the other hand, by increasing $t_{\text{S}}$, the two F layers tend to decouple 
and the current through the structure is only due to quasiparticles flowing 
separately through them, independently of $t_{\text{S}}$.
The resulting MR ratio is shown in Fig.~\ref{tS}(b) and exhibits very large 
\emph{negative} values around $-70\%$ for $t_{\text{S}}\simeq 3.5\xi_{0}$ and 
 about $-25\%$ for $t_{\text{S}}\simeq 6.5\xi_{0}$. It is noteworthy to mention that when the S layer is in the normal state (i.e., a FNF trilayer) 
 $\text{MR}\simeq (0.7\pm 1.8)\%$ for $t_{\text{S}}=4.5\xi_{0}$.

The role the F layers thickness
on the conductance and magnetoresistance is analyzed in Fig. \ref{tF}, for fixed $t_{\text{S}}=40$ and $\mathcal{P}=100\%$. 
Figure \ref{tF}(a) shows that the conductance in the P alignment increases 
linearly with $t_{\text{F}}$ according to the estimate in Eq.~\ref{gp}.
In the AP configuration the conductance is again linear in $t_{\text{F}}$ with the same slope, but it is shifted upwards as compared to $G_{\text{P}}$.
This is in agreement with Eq.~(\ref{gap}): the difference $G_{\text{P}}-G_{\text{AP}} \sim G^{\ast}$.
As a consequence, the MR ratio (see Fig. \ref{tF}(b)) starts from $\simeq-70\%$ 
at $t_{\text{F}}\simeq 0.5\xi_{0}$ and thereafter decreases by increasing the value of $t_{\text{F}}$.

\begin{figure}[t!]
\includegraphics[width=\columnwidth,clip]{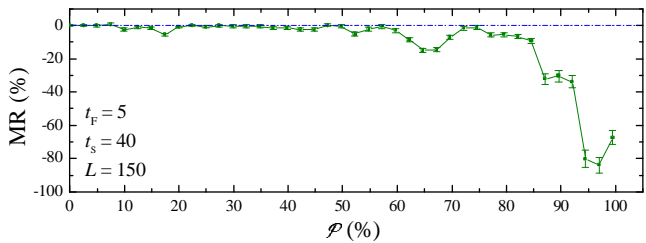}
\caption{ (color online) MR ratio versus $\mathcal{P}$ with $t_{\text{S}}=40$, 
	and $t_{\text{F}}=5$. The same parameters as for Fig.~\ref{tS} were used.}
\label{exch}
\end{figure}

We finally analyze the behavior of MR, for $t_{\text{S}}=40$ and $t_{\text{F}}=5$, as a function of the polarization of the F layers. Figure~\ref{exch} shows that the value of the MR ratio remains smaller than $\sim -30\%$ up to $\mathcal{P}\simeq 87\%$ and then grows to larger negative values. Highly spin-polarized ferromagnets are thus required for the effect to be maximized. 
The fluctuations present in the MR$(\mathcal{P})$ curve 
can be ascribed to opening and closing of conducting channels in the F layers as well as to size effects.

A 3D structure was also considered, allowing the system to 
extend in the $z$ direction (see Fig. 1). The calculations, performed for 
several values of the structure width ($w$), confirmed qualitatively the overall results found in the 2D case. 
We finally stress the importance of a good metallic contact between F and S layers. The presence of a barrier at the FS interface would indeed lead to a suppression of CAR and therefore of the MR value.

In conclusion, we have investigated theoretically spin transport in a ferromagnet-superconductor-ferromagnet trilayer in the current-in-plane geometry.  We showed that very large and negative magnetoresistance 
values (exceeding $-80\%$) can be achieved. Such an effect relies entirely on the existence of crossed Andreev reflection. The  results presented here are 
promising in light of the implementation of novel-concept magnetoresistive 
devices such as, for instance, spin-switches as well as magnetoresistive 
memory elements. To this end, half-metallic ferromagnets such as 
CrO$_2$~\cite{coey,CrO2a}, NiMnSb \cite{NiMnSb}, Sr$_2$FeMoO$_6$~\cite{SrFeMoO} 
and La$_{2/3}$Sr$_{1/3}$MnO$_3$~\cite{bowen} appear as particularly suitable. 
Also the Ga/Ga$_{1-x}$Mn$_x$As material system, exploiting a superconductor 
in combination with heavily-doped ferromagnetic semiconductor layers, 
appears to be a good candidate for the implementation of this structure, 
thanks to Ga$_{1-x}$Mn$_x$As predicted half-metallic nature (for $x\geq 0.125$)~\cite{ogawa} and to its well-developed technology~\cite{braden}.

We thank V. Dediu, M. V. Feigel'man, and S. Sanvito for valuable discussions.
This work was partially supported by MIUR under FIRB "Nanotechnologies and 
Nanodevices for Information Society", contract RBNE01FSWY.


\begin{thebibliography}{99}
\bibitem{AFM}
	Usually, an additional insulating antiferromagnetic 
	layer is placed below F$_2$ in order to pin its (in-plane) direction 
	of magnetization via the exchange bias, so that it remains fixed in 
	weak magnetic fields that can orient the magnetization of the "free" 
	layer F$_1$ \cite{wolf01}.
\bibitem{GTV}
	Finite temperature and finite voltage differential conductance can be evaluated using the theory developed  in Ref. \cite{lambert93}.
\bibitem{ARCAR}
	Note that $R_{\text{a}}^{\sigma}$ includes both AR and CAR processes. Quasiparticles injected and arriving at the same F layer are accounted for by AR, while quasiparticles injected and arriving at different F layers are accounted for by CAR. In the half-metallic case, however, $R_{\text{a}}^{\sigma}$ describes only CAR, since AR processes are totally hindered.
\bibitem{self}
	We expect that for thin ($t_\text{S}/\xi_0\leq 1$) S layers the MR effect is quite reduced due both to a gap suppression \cite{you04} and to direct quasiparticle tunneling which becomes dominant through thinner superconducting layers.
\bibitem{polarizzazione}
	The ferromagnet bulk polarization is defined as $\mathcal{P}=	
	h_{\text{exc}}/(4\gamma-\epsilon_{\text{S}})$.
\end{thebibliography}
\end{document}